\documentclass{ieeeaccess}
\usepackage{cite}
\usepackage{amsmath,amssymb,amsfonts}
\usepackage{graphicx}
\usepackage{textcomp}

\usepackage{longtable}
\usepackage{mathtools}
\usepackage{empheq}
\mathtoolsset{showonlyrefs}
\usepackage[acronym]{glossaries}
\usepackage{acronym}
\usepackage{algpseudocode, algorithm}
\usepackage{indentfirst}
\usepackage{amsthm}
\usepackage{amsfonts}
\usepackage{multirow}
\usepackage{color}
\usepackage{MnSymbol}
\usepackage{subfig}
\usepackage{fmtcount} 
\usepackage{enumitem}
\usepackage[open,openlevel=1]{bookmark}
\usepackage{xcolor}
\usepackage{balance}

\usepackage{url}

\usepackage{breakurl}

\usepackage{bm}
\usepackage{hyperref}
\usepackage{makecell}
\usepackage{amsbsy}

\def\BibTeX{{\rm B\kern-.05em{\sc i\kern-.025em b}\kern-.08em
    T\kern-.1667em\lower.7ex\hbox{E}\kern-.125emX}}
\begin{document}
\history{Date of publication xxxx 00, 0000, date of current version xxxx 00, 0000.}
\doi{10.1109/ACCESS.2020.DOI}

\title{Artificial Intelligence (AI) and Big Data for Coronavirus (COVID-19) Pandemic: \\A Survey on the State-of-the-Arts}
\author{\uppercase{Quoc-Viet Pham}\authorrefmark{1},
\uppercase{
	Dinh C. Nguyen\authorrefmark{2}, Thien Huynh-The\authorrefmark{3},\\
	Won-Joo Hwang\authorrefmark{4,5}, and Pubudu N. Pathirana}\authorrefmark{2}
}
\address[1]{Research Institute of Computer, Information and Communication, Pusan National University, Busan 46241,  Korea (e-mail: vietpq@pusan.ac.kr).}
\address[2]{School of Engineering, Deakin University, Waurn Ponds, VIC 3216, Australia (e-mail: \{cdnguyen, pubudu.pathirana\}@deakin.edu.au).}
\address[3]{ICT Convergence Research Center, Kumoh National Institute of Technology, Gyeongsangbuk-do 39177, Korea (e-mail: thienht@kumoh.ac.kr).}
\address[4]{Department of Biomedical Convergence Engineering, Pusan National University, Busan, 46241 Korea.}
\address[5]{Department of Information Convergence Engineering (Artificial Intelligence), Pusan National University, Busan 46241, Korea (e-mail: wjhwang@pusan.ac.kr).}
\tfootnote{This work was supported by the National Research Foundation of Korea (NRF) Grant funded by the Korea Government (MSIT) under Grants NRF-2019R1C1C1006143 and NRF-2019R1I1A3A01060518. This work was also supported by Institute of Information \&
communications Technology Planning \& Evaluation (IITP) grant funded by the Korea government (MSIT) (No. 2020-0-01450, Artificial Intelligence Convergence Research Center [Pusan National University]).}

\markboth{IEEE Access}{Q.-V. Pham \headeretal: Artificial Intelligence and Big Data for Coronavirus Pandemic}

\corresp{Corresponding author: Won-Joo Hwang (e-mail: wjhwang@pusan.ac.kr).}

\begin{abstract}
The very first infected novel coronavirus case (COVID-19) was found in Hubei, China in Dec. 2019. The COVID-19 pandemic has spread over 214 countries and areas in the world, and has significantly affected every aspect of our daily lives. At the time of writing this article, the numbers of infected cases and deaths still increase significantly and have no sign of a well-controlled situation, e.g., \textcolor{black}{as of 13 July 2020, from a total number of around 13.1 million positive cases, $571,527$ deaths} were reported in the world. Motivated by recent advances and applications of artificial intelligence (AI) and big data in various areas, this paper aims at emphasizing their importance in responding to the COVID-19 outbreak and preventing the severe effects of the COVID-19 pandemic. We firstly present an overview of AI and big data, then identify the applications aimed at fighting against COVID-19, next highlight challenges and issues associated with state-of-the-art solutions, and finally come up with recommendations for the communications to effectively control the COVID-19 situation. It is expected that this paper provides researchers and communities with new insights into the ways AI and big data improve the COVID-19 situation, and drives further studies in stopping the COVID-19 outbreak.
\end{abstract}


\titlepgskip=-15pt

\maketitle

\section{Introduction}
\label{Sec:Introduction}
\textcolor{black}{Coronavirus disease-19 (COVID-19), caused by a novel coronavirus, has changed the world significantly, not only in the health care space, but also in many aspects of human life such as education, transportation, politics, supply chain, etc.} Infected COVID-19 people normally experience respiratory illness and can recover with effective and appropriate treatment methods. What makes COVID-19 much more dangerous and easily spread than other Coronavirus families is that the COVID-19 coronavirus has become highly efficient in human-to-human transmissions. As the writing of this paper, the COVID-19 virus has spread rapidly in $ 214 $ countries.
The United State of America (USA) 
currently records with the highest COVID-19 cases, \textcolor{black}{more than $ 3.4 $ million confirmed cases and nearly $ 137,782 $ deaths (July 13, 2020)}. Some other countries like Brazil, India, Russia, and Spain are also enormously influenced. However, there are no clinical vaccines to prevent the COVID-19 virus and specific drugs/therapeutic protocols to combat this communicable disease. 


As the leaders in the war against the novel coronavirus, the World Health Organization (WHO) and Centers for Disease Control and Prevention (CDC) have released a set of public advices and technical guidelines \cite{WHO_COVID,CDC_COVID}. The cooperation between and efforts from national governments and large corporations are expected to significantly reduce risks from the spread of COVID-19 outbreak. For example, as a search engine giant, Google launched a COVID-19 portal (\url{www.google.com/covid19}), where we can find useful information, such as coronavirus map, latest statistics, and most common questions on COVID-19. Another example is that IBM, Amazon, Google, and Microsoft with White House developed a supercomputing system for researches relevant to the coronavirus
\cite{White_House}. 
In response to the pandemic, some publishers now offer free access to the articles, technical standards, and other documents related to the COVID-19-like virus, while web archival services like arXiv, medRxiv, and bioRxiv create a fast link to collected all preprints related to COVID-19 \cite{arXiv2020}. 
On the other hand, artificial intelligence (AI) and big data have found in a lot of applications in various fields, e.g., AI in computer science, AI in banking, AI in agriculture, and AI in healthcare. These technologies are expected  and may play important roles in the global battle against the COVID-19 pandemic. 

To better understand and alleviate the COVID-19 pandemic, many papers and preprints have been published online in the last few months. Our main purpose is to show the effectiveness of AI and big data to fight against the COVID-19 pandemic and review state-of-the-art solutions using these technologies. Moreover, two representative examples of the AI and big data-based solutions are presented for a better understanding. Besides, we highlight challenges and issues associated with existing AI and big data-based approaches, which motivate us to produce a set of recommendations for the research communities, governments, and societies. \textcolor{black}{We note that the papers to be reviewed in this work are collected from various sources like IEEE Xplore, Nature, ScienceDirect, and Wiley by using a subsequent of strings ("AI" OR "artificial intelligence") AND ("big COVID-19" OR "Coronavirus"  OR "SARS-CoV-2") or ("big data" OR "data science") AND ("big COVID-19" OR "Coronavirus" OR "SARS-CoV-2"). Moreover, as a large number of preprints have been published online since Dec. 2019, representative preprints from archived websites such as arXiv, medRxiv, and bioRxiv are also selected.}

This work is organized as follows. Section~\ref{Sec:Fundamentals} presents fundamental knowledge of COVID-19, AI, big data, and shows primary motivations behind the use of AI and big data.
Next, applications of AI and big data in fighting the COVID-19 pandemic, e.g., identifying the infected patients, tracking the COVID-19 outbreak, developing drug researches, and improving the medical treatment, are reviewed and summarized in Sections~\ref{Sec:Applications_AI} and~\ref{Sec:Applications_BD}, respectively. 
In Section~\ref{Sec:Examples}, two examples of AI and big data-based frameworks for combating the COVID-19 disease are described: a phone-based framework for the COVID-19 detection/surveillance, and a data-driven approach using AI for finding antibody sequences.
Section~\ref{Sec:Lessons} highlights lessons and recommendations learned from this paper, and discusses the challenges needed to be solved in the future. Finally, Section~\ref{Sec:Conclusion} concludes this paper. 


\section{Background and Motivations}
This section presents the fundamentals of COVID-19, AI, big data. This section also shows motivations behind the use of AI and big data in response to the COVID-19 pandemic. 
\label{Sec:Fundamentals}

\begin{figure*}[t]
	\centering
	\includegraphics[width=0.975\linewidth]{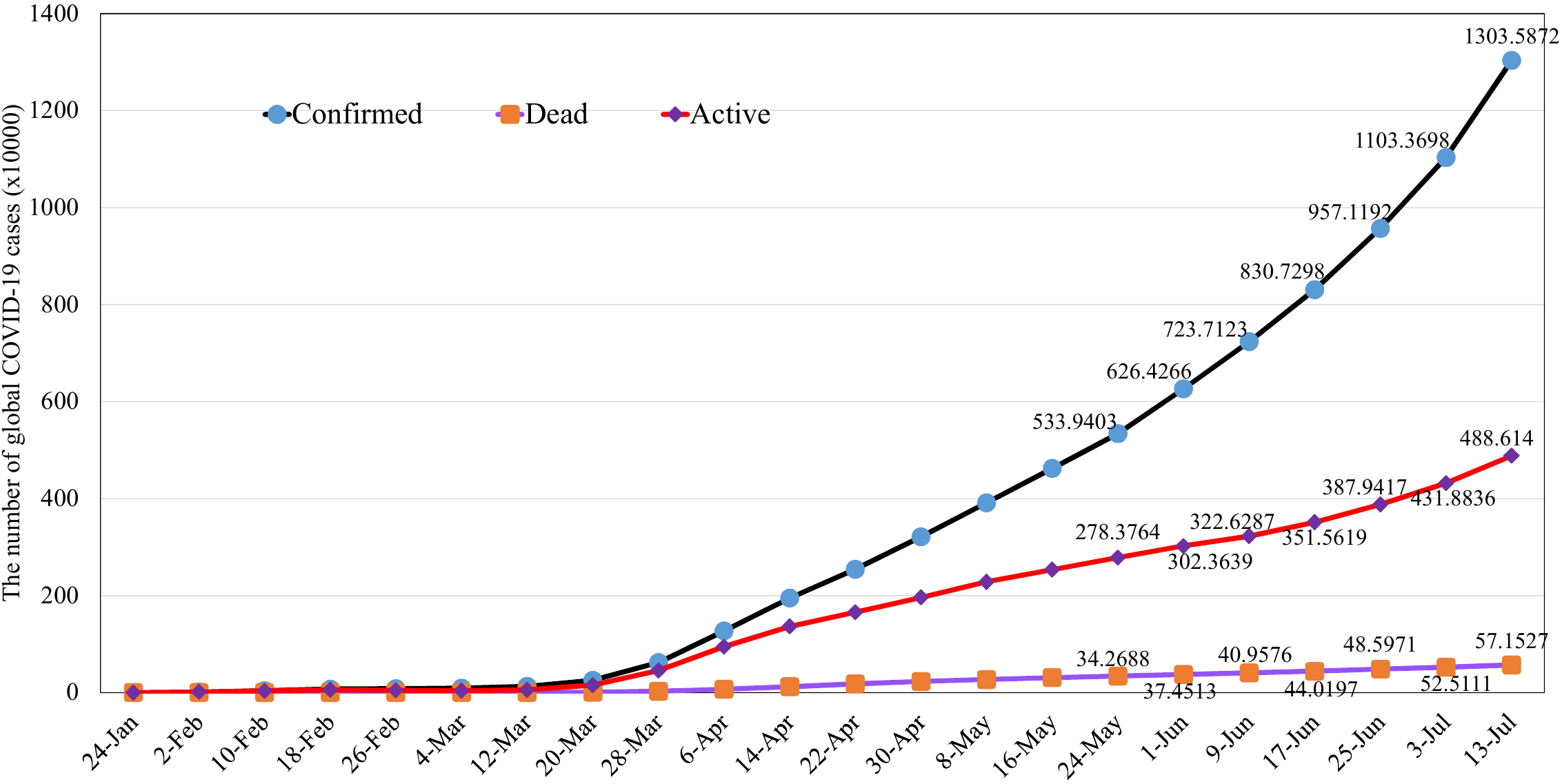}
	\caption{The global COVID-19 trend (Source: CoronaBoard). \textcolor{black}{Data accurate as of July 13, 2020}.}
	\label{Fig:COVID19_Trend}
\end{figure*}

\subsection{COVID-19 pandemic}
COVID-19 is caused by severe acute respiratory syndrome coronavirus 2 (SARS-CoV-2), a betacoronavirus \cite{SOHRABI202071}. The first COVID-19 infected case was reported in Wuhan City, Hubei Province of China on 31 December 2019, and has quickly spread out almost all countries in the world, $ 214 $ countries and areas at the time of writing this article. There is no sign that the numbers of infected and dead cases will decrease and the situation will be under control.
\textcolor{black}{
In particular, the numbers of new cases are still very large, around $ +139,000 $ and $ +3,000 $ infected and dead cases per day, as report by the CoronaBoard\footnote{https://coronaboard.kr/}, whereas the fatality rate is $ 4.38 $\% (accurate as of July 13 2020).} 
The global COVID-19 trend can be found in Fig.~\ref{Fig:COVID19_Trend}. Due to the serious situation of COVID-19, WHO has increased the COVID-19 risk assessment to the highest level and declared it as a global pandemic. For a better understanding of the COVID-19 virus about its structure, etiology and pathogenesis, characteristics, and recent treatment progress, we refer the interested readers to refer the paper \cite{Li2020Coronavirus}.

As the dramatic impact of COVID-19 on the globe, lots of attempts have been paid for solutions to fight against the COVID-19 outbreak. Government's efforts are mainly responsible to stop the pandemic, e.g., lock down the (partial) area to limit the spread of infection, ensure that the healthcare system is able to cope with the outbreak and provide crisis package to alleviate impacts on the national economics and people, and adopt adaptive policies according to the COVID-19 situation. At the same time, individuals are encouraged to stay healthy and protect others by following some advices like wearing the mask at public locations, washing the hands frequently, maintaining the social distancing policy, and reporting the latest symptom information to the regional health center\footnote{https://www.who.int/health-topics/coronavirus}. On the other hand, research and development relevant to COVID-19 are now prioritized, and have been received special interest from various stakeholders like governments, industries, and academia. 
\textcolor{black}{For example, studies in \cite{ivanov2020viability, ivanov2020predicting} showed huge impacts of the COVID-19 pandemic on the global supply chain, and considered different aspects of supply chains, including viability, stability, robustness, and resilience.}
Besides the global attempt to develop an effective vaccine and medical treatment for the COVID-19 coronavirus, computer science researchers made initial efforts for the fight against COVID-19. Motivated by the tremendous success of AI and big data in various areas, we present state-of-the-art solutions and approaches based on AI and big data for tackling the COVID-19 coronavirus disease.

\subsection{Artificial Intelligence}
AI is a thriving technology for many intelligent applications in various fields. Some high-profile examples of AI are autonomous vehicles (e.g., self-driving car and drones) in automotive, medical diagnosis and telehealth in healthcare, cybersecurity systems (e.g., malware and botnet detection), AI banking in finance, image processing and natural language processing in computer vision. Among many branches of AI, machine learning (ML) and deep learning (DL) are two important approaches. Generally, ML refers to the ability to learn and extract meaningful patterns from the data, and the performance of ML-based algorithms and systems are heavily dependent on the representative features. In the meanwhile, DL is able to solve complex systems by learning from simple representations. According to \cite{goodfellow2016deep, 8963964}, DL has two main features: 1) the ability to learn the right representations shows one feature of DL and 2) DL allows the system to learn the data from a deep manner, multiple layers are used sequentially to learn increasingly meaningful representations.

AI offers a powerful tool to fight against the COVID-19 pandemic \cite{bragazzi2020big}. For example, the scientists in \cite{Beck2020Predicting} developed a DL model to identify existing and commercial drugs for the ``drug-repurposing" (also known as drug repositioning), i.e., finding a rapid drug strategy using existing drugs that can be immediately applied to the infected patients. This study is motivated by the fact that newly developed drugs usually take years to be successfully tested before coming to the market. Although findings in this study are currently not clinically approved, they still open new ways to fight the COVID-19 disease. \textcolor{black}{The work in \cite{Zhavoronkov2020Potential}} proposed using the deep generative model for drug discovery (which is defined as the process of identifying new medicines). The COVID-19 protease structures generated from the DL model in this work would be further used for computer modeling and simulations for the purpose of obtaining new molecular entity compounds against the COVID-19 coronavirus. 
Utilizing DL for computed tomography (CT) image processing, the authors in \cite{Zheng2020DeepLearning} showed that their proposed DL model when training with $ 499 $ CT volumes and testing on $ 131 $ CT volumes, can achieve an accuracy of $ 0.901 $ with a positive predictive value of 0.840 and a negative predictive value of $ 0.982 $. This study offers a fast approach to identify the infected COVID-19 patient, which may provide great helps in timely quarantine and medical treatment. 
\textcolor{black}{The last example is the use of AI (i.e., an autoencoder based method) in real-time predicting dynamics (e.g., the number of infected cases, ending time and trajectory of the COVID-19 pandemic) of the COVID-19 outbreak in China \cite{hu2020artificial}.} The high accuracy of the AI-based approach proposed in \cite{hu2020artificial} is helpful in monitoring the COVID-19 outbreak and improving the health and policy strategies.
Along with the applications mentioned above, the involvement of giant techs is needed because researchers, doctors and scientists can be effectively supported to expedite the research and development of COVID-19 virus. 
Recently, IBM announced that they are now providing a cloud-based research resource that has been trained on \textcolor{black}{a COVID-19 open dataset (CORD-19) \cite{CORD-19}, which is a collection of research articles related to COVID-19.} 
Moreover, IBM has adopted their proposed AI technology for drug discovery, from which 3000 novel COVID-19 molecules have been obtained, officially reported in \cite{IBMReleases2020}. Another support is from the White House Office of Science and Technology Policy, the U.S. Department of Energy and IBM with the development of COVID-19 HPC Consortium (\url{https://covid19-hpc-consortium.org/}), which is open for research proposals concerning COVID-19. 
\textcolor{black}{Another example is the Coronavirus International Research Team (COV-IRT) (\url{https://www.cov-irt.org/}), a group of scientists who are developing vaccines and therapeutic solutions against COVID-19.}

\subsection{Big Data}
\subsubsection{Definition and Characteristics} The rapid development of the Internet of Things (IoT) results in a massive explosion of data generated from ubiquitous wearable devices and sensors \cite{8985278}. The unprecedented increase of data volumes associated with advances of analytic techniques empowered from AI has led to the emergence of a big data era 
\cite{1, 8985278}.
Big data has been employed in a wide range of industrial application domains, including healthcare where electronic healthcare records (EHRs) are exploited by using intelligent analytics for facilitating medical services. For example, health big data potentially supports patient health analysis, diagnosis assistance, and drug manufacturing \cite{3}.
Big data can be generated from a number of sources which may include online social graphs, mobile devices (i.e. smartphones), IoT devices (i.e. sensors), and public data \cite{5} in various formats such as text or video. In the context of COVID-19, big  data  refers  to  the  patient  care  data such as physician notes, X-Ray reports, case  history,  list  of  doctors  and  nurses, and information of outbreak areas. \textcolor{black}{In general, big data is the information asset characterized by such a high volume, velocity and variety to acquire specific technology and analytical methods for its transformation into useful information to serve the end users, i.e, big data in digital twin technologies \cite{twin1,twin2,twin3,twin4}}. The three characteristics of big data are summarized as follows. 

\begin{itemize}
	\item 	\textit{Volume:} This feature shows the huge amount of data that can range from terabytes to exabytes. According to a Cisco's forecast, the data traffic is expected to reach $ 930 $ exabytes by 2020, a seven-fold growth from 2017 \cite{6}.
	\item 	\textit{Variety:} It refers to the diversity and heterogeneity of big data. For example, big data in healthcare can be produced from healthcare users (i.e. doctors, patients), medical IoT devices, and healthcare organizations. Data can be formatted in text, images, videos with structured or un-structured dataset types \cite{7}. 
	\item 	\textit{Velocity:} It expresses the data generation rate that can be calculated in time or frequency domain. In fact, in industrial applications like healthcare, data generated from devices is always updated in real-time, which is of significant importance for time-sensitive applications such as health monitoring or diagnosis \cite{8}. 
\end{itemize}

\subsubsection{Big data for COVID-19 fighting}
Big data has been proved its capability to support fighting infectious diseases like COVID-19 \cite{9}, \cite{10}. Big data potentially provide a number of promising solutions to help combat COVID-19 epidemic. By combining with AI analytics, big data helps us to understand the COVID-19 in terms of outbreak tracking, virus structure, disease treatment, and vaccine manufacturing \cite{11}. For example, big data associated with intelligent AI-based tools can build complex simulation models using coronavirus data streams for outbreak estimation. This would aid health agencies in monitoring the coronavirus spread and preparing better preventive measurements \cite{12}. Models from big data also supports future prediction of COVID-19 epidemic by its data aggregation ability to leverage large amounts of data for early detection. Moreover, the big data analytics from a variety of real-world sources including infected patients can help implement large-scale COVID-19 investigations to develop comprehensive treatment solutions with high reliability \cite{13}, \cite{14}. This would also help healthcare providers to understand the virus development for better response to the various treatment and diagnoses. 

Based on the above analysis, we want to highlight that big data analytics is the process of collecting and analyzing the large volume of data sets to discover useful hidden patterns and other information, e.g., COVID-19 data discovering. \textcolor{black}{Moreover, AI (and explainable AI \cite{ExplainableAI}) aims to apply logic and reasoning to build human intelligence that can mimic the function of a machine for learning, classifying, and estimating possible outcomes, e.g., COVID-19 symptom classifications.} The potential applications of each technology in fighting COVID-19 pandemic will be explained and discussed in the following sections through a number of practical use cases.

%
\section{Applications of AI for Fighting Covid-19}
This section presents representative applications of AI in fighting the Covid-19 outbreak. 
\label{Sec:Applications_AI}
\subsection{AI for COVID-19 detection and diagnosis}
As one of the most effective solutions to combat the COVID-19 pandemic, early treatment and prediction are of importance. Currently, the standard method for classifying respiratory viruses is the reverse transcription polymerase chain reaction (RT-PCR) detection technique. In response to the COVID-19 virus, some efforts have been dedicated to improve this technique \cite{corman2020detection} and for other alternatives \cite{fomsgaard2020alternative}. These techniques are, however, usually costly and time-consuming, have low true positive rate, and require specific materials, equipment and instruments. Moreover, most countries are suffering from a lack of testing kits due to the limitation on budget and techniques. Thus, the standard method is not suitable to meet the requirements of fast detection and tracking during the COVID-19 pandemic. A simple and low-cost solution for COVID-19 identification is using smart devices together with AI frameworks \cite{maghdid2020novel, rao_vazquez_2020}. This is referred to as \emph{mobile health} or \emph{mHealth} in the literature \cite{SILVA2015265}. These works are advantageous since smart devices
are daily used for multi-purposes. Moreover, the emergence of cloud and edge computing can effectively overcome the limitation of batter, storage, and computing capabilities \cite{Pham2019ASurvey_MEC}.

\begin{figure*}[t]
	\centering
	\includegraphics[width=0.65\linewidth]{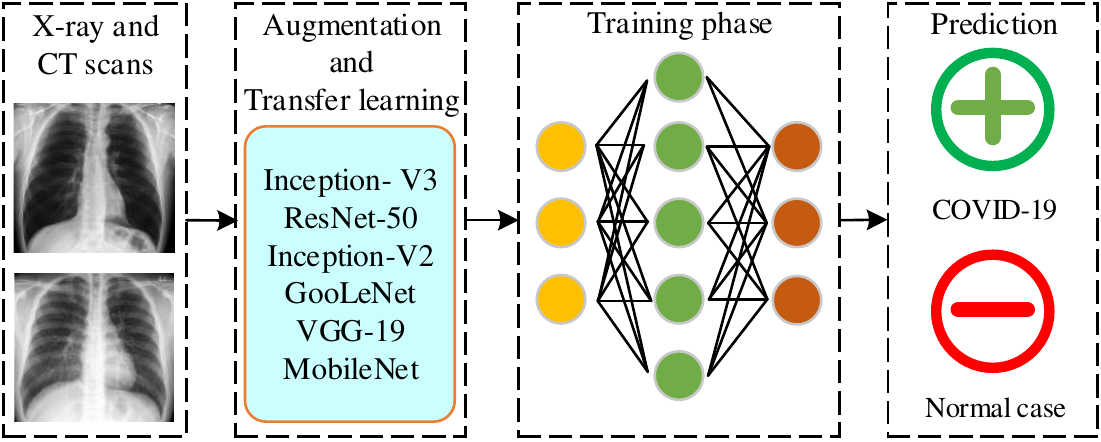}
	\caption{An illustration of DL-based frameworks for the COVID-19 detection and diagnosis.}
	\label{Fig:AI_MedicalImage_Model}
\end{figure*}

Another directive for COVID-19 detection is to use AI techniques for medical image processing, which recently appeared in many research works on coronavirus 
\cite{gozes2020coronavirus, barstugan2020coronavirus, PANWAR2020109944, khalifa2020detection, abbas2020classification, asnaoui2020automated, apostolopoulos2020covid, narin2020automatic, UCAR2020109761, ghoshal2020estimating}.
As we limit this paper to the COVID-19 virus, the interested readers are invited to read the surveys in \cite{litjens2017survey, shen2017deep} for other applications of DL in medical image analysis. It is noted from these works that X-ray images and computed tomography (CT) scans are widely used as the input of DL models so as to automatically detect the infected COVID-19 case. Motivated by an important finding that infected COVID-19 patients normally reveal abnormalities in chest radiography images, the authors in \cite{wang2020covid} designed a deep convolutional neural network (CNN) model for the detection of COVID-19 cases. The three-class classification (normal, COVID-19 infected, and non-COVID-19 infected) in this work is helpful if the medical staff needs to decide which cases should be tested with standard methods (between normal and COVID-19 infected cases), and which treatment strategies should be taken (between non- and COVID-19 infected cases). By training on an open source \textcolor{black}{dataset with $ 13,975 $ images of $ 13,870 $ patients}, the proposed CNN model can achieve \textcolor{black}{an accuracy of $ 93.3 $\%}. The use of ML and DL techniques with chest CT scans for COVID-19 detection were considered in \cite{gozes2020coronavirus, gozes2020rapid, barstugan2020coronavirus, wang2020deep, ozkaya2020coronavirus, Wei0846537120913033}, respectively. These works show a high performance as they can achieve a high classification accuracy, e.g., $ 99.68 $\% in \cite{gozes2020coronavirus}, an area under curve (AUC) score of $ 0.994 $ in \cite{barstugan2020coronavirus}, AUC of $ 0.996 $ in \cite{gozes2020rapid}, and accuracy of $ 82.9 $\% ($ 98.27 $\%) with specificity of $ 80.5 $\% ($ 97.60 $\%) and sensitivity of $ 84 $\% ($ 98.93 $\%) in \cite{wang2020deep} and \cite{ozkaya2020coronavirus}. As the cost of X-ray scans is usually cheaper than CT images, a large portion of research works for COVID-19 detection utilize DL models with \textcolor{black}{X-ray} images. For example, the work in 
\cite{abbas2020classification}
leveraged a deep CNN model, called Decompose, Transfer, and Compose (DeTraC), to process chest X-ray images for the classification of COVID-19. The main purpose of the decomposition layer is to reduce the feature space, thus yielding more sub-classes but improving the training efficiency, whereas the composition layer is to combine sub-classes from the previous layer so as to produce the final classification result. 
Besides the decomposition and composition layers, a transfer layer is positioned in the middle to speed up the training time, reduce the computation costs, and make the DL model trainable with small datasets. The importance of transfer learning made it a wide consideration in many works, e.g., \cite{narin2020automatic, khalifa2020detection, apostolopoulos2020covid}. To summarize this paragraph, a general representation of DL-based models for the COVID-19 detection and diagnosis can be showed in Fig.~\ref{Fig:AI_MedicalImage_Model}, where we present the output as a binary classification: COVID-19 infected and normal. 

In the fight against the COVID-19 pandemic, developing efficient diagnostic and treatment methods plays an important role in mitigating the impact of COVID-19 virus. The work in \cite{chaganti2020quantification} introduces a method based on DL and deep reinforcement learning (DL) to quantify abnormalities in the COVID-19 disease. The input into the proposed learning model is non-contrasted chest CT images, while the output is severity scores, including percentage of opacity (POO), lung severity score (LSS), percentage of high opacity (POHO), and lung high opacity score (LHOS). The proposed learning model, when trained and tested on a dataset of $ 568 $ CT images and $ 100 $ samples respectively, shows promising results as the Pearson's correlation coefficient between the ground truth and predicted output is $ 0.97 $ for POO, $ 0.98 $ for POHO, $ 0.96 $ for LSS, and $ 0.96 $ for LHOS.
Another work leveraging DL with CT images is in \cite{shan+2020lung}, where two measurement metrics: volume of infection and percentage of infection are quantified. 
In contrast with other studies, the work in \cite{ghoshal2020estimating} exploited a pre-trained ResNet50V2 model \cite{he2016identity} with the Bayesian DL classifier to estimate the diagnostic uncertainty.   
A random forest (RF) model was investigated in \cite{tang2020severity} to assess the COVID-19 severity, where $ 63 $ features, e.g., infection ratio of the lung and the ground class opacity lesions, are extracted from chest CT images. The result is very promising as it can achieve sensitivity of $ 0.933 $, selectivity of $ 0.745 $, accuracy of $ 0.875 $, and AUC score of $ 0.91 $. The key finding from this study is that the severity level is more dependent on the features extracted from the right lung.  
As an effort to develop a scalable and cost-effective solution for COVID-19 diagnosis, the authors in \cite{imran2020ai4covid19} proposed an AI-based framework called AI4COVID-19, which takes into account the deep domain knowledge of medical experts, and uses smart phones to record cough/sound signals as the input data. In particular, the method relies on the fact that COVID-19 disease is likely to have
idiosyncrasies from the underlying pathomorphological alterations, e.g., ground-glass opacity ($ 91 $\% for COVID-19 versus $ 68 $\% for non-COVID-19), vascular thickening ($ 59 $\% for COVID-19 versus $ 22 $\% for non-COVID-19). Despite interesting results (e.g., \textcolor{black}{average accuracy of $ 97.91 $\% (93.56\%) for cough (COVID-19) detection}, and response time within $ 60 $ seconds), the implementation of AI4COVID-19 at a larger scale is currently limited by several issues, such as quantity and quality of the datasets, and lack of clinical validation. 


\subsection{Identifying, tracking, and predicting the outbreak}


A traditional compartmental model to predict the infectious disease is the SIR (Susceptible-Infected-Removed) model. This SIR model can be 
expressed by the following non-linear differential equations \cite{Zhong2020Sharp}: 
\begin{align}
dS / dt = -\beta S I, \;\;
dI / dt = \beta S I - \gamma I, \;\;
dR / dt = -\gamma I,
\end{align}
where $ S $, $ I $, and $ R $ denote susceptible, infected, and removed individuals, respectively, $ \beta $ is the transmission rate, and $ \gamma $ is the recovering rate. However, this model is not suitable for the COVID-19 pandemic because of its assumptions, which are 1) the recovered cases will not get infected again and 2) the model simply ignores the time-varying nature of two parameters $ \beta $ and $ \gamma $. To overcome issues associated with the traditional SIR model, some efforts have been paid for properly adapting the COVID-19 pandemic. For example, a time-dependent SIR model was proposed in \cite{chen2020time}, which can adapt to the change of infectious disease control and prevention laws such as city lock downs and traffic halt, as the control parameters $ \beta $ and $ \gamma $ are modeled as time-variant variables. This model can be further improved by having two types of infected patients: detectable and undetectable, which the former has a lower transmission rate than that of the latter. Numerical results in \cite{chen2020time} show the proposed time-dependent model can accurately predict the number of confirmed cases in China, and the outbreak time of some countries like Republic of Korea, Italy, and Iran. Several variants of the SIR model other studies, e.g., space-time dependence of the COVID-19 pandemic in \cite{biswas2020space}, SIQR model in \cite{crokidakis2020data}, where an additional compartment, namely Quarantined, is considered in the traditional SIR model, and stochastic SIR model in \cite{gaeta2020simple}. As a summary, these models can predict the outbreak of COVID-19 and send alarm messages to the governments and policy makers so that some actions should be taken in advance prior to the outbreak. 


To model the outbreak size of COVID-19, over the last few months, a number of research works have been done using ML and DL models. For example, the authors in \cite{dandekar2020neural} argued that the traditional SIR model is not able to capture the effects of more granular interactions such as social distancing and quarantine policies. The authors proposed encoding the quarantine policy as a strength function, which is then included in a neural network for predicting the outbreak size in in
Wuhan, China. The experimental results using the publicly available data from the Chinese National Health Commission
demonstrate that the quarantine policy plays an important role in controlling the COVID-19 outbreak, and the number of infected cases can increase exponentially without the right quarantine policy. Applying the proposed model with quarantine for the United State of America (USA), the infection of COVID-19 may show a stagnation by 20 April if the current COVID-19 policies are executed without any changes. The authors also applied their proposed learning model in \cite{dandekar2020neuralglobal} to estimate the global outbreak size and obtained similar results. Very recently, some scholars proposed working directly with the empirical data to estimate the outbreak size, for example, logistic model in \cite{Zhou2020Forecasting}, Bayesian non-linear model in \cite{bayes2020modelling}, Prophet forecasting procedure in \cite{ndiaye2020analysis}, and autoregressive integrated moving average (ARIMA) model in \cite{perone2020arima}. The study in \cite{magdon2020machine} presents a data-driven approach to learn the phenomenology of COVID-19. The proposed model enables us to read asymptomatic information, e.g., a lag (also known as  incubation period) of about $ 10 $ days and virulence of $ 0.14 $\%. 

Another interesting work in \cite{hu2020forecasting}  proposed a modified auto-encoder (MAE) for modeling the transmission dynamics of COVID-19 in the world. Using the data collected from the WHO situation reports, it is shown that the proposed MAE model can predict the outbreak size accurately with an average error less than $ 2.5 $\%. The experimental results also show that the COVID-19 outbreak can be prevented effectively with fast public health intervention, e.g., compared with one-month intervention, one-week case can reduce the maximum number of cumulative and dead
cases by around 166.89 times. Applying the MAE model for the surveillance data of the confirmed Covid-19 cases in China \cite{hu2020artificial}, the authors also showed high forecasting capabilities of the proposed MAE model for the transmission dynamics and plateau of COVID-19 in China. The authors in \cite{pal2020neural} proposed combining medical information (e.g., susceptible and dead cases) and local weather data to predict the risk level of the country. Specifically, a shallow long short-term memory (LSTM) neural network is used to overcome the challenges of small dataset, and the risk level (high, medium, and recovering) of a country is classified by using the fuzzy rule. The experiment results show that the proposed learning model can achieve an average accuracy of $ 78 $\% over 170 countries. 

\subsection{AI for Infodemiology and Infoveillance}
To date, most reliable information on the COVID-19 pandemic is disseminated through the official websites and social channels of the health organizations like the WHO, and the ministry of health and welfare in each country. However, electronic medium and online platforms (e.g., Facebook, Twitter, YouTube, and Instagram) have showed their importance in distributing information related to COVID-19 like pandemic. Regardless of the quality and source, information from the media platform and the Internet is highly accessible and timely, so further analyses can be performed if the data can be collected and processed properly. As a powerful tool to deal with a vast amount of data, AI have been utilized to have a better understanding of the social network dynamics and to improve the COVID-19 situation. To illustrate the applications of AI during the disease pandemic, the work in \cite{Ganasegeran2020} presented some real-life examples: 1) using Twitter data to track the public behavior, 2) examining the health-seeking behavior of the Ebola outbreak, and 3) public reaction towards the Chikungunya outbreak.  

Similar to the previous pandemics, the recent emergence of COVID-19 has called for several studies from the infodemiology and infoveillance perspectives. The authors in \cite{hou2020assessment} analyzed the data collected from three popular social platforms in China: Sina Weibo, Baidu search engine, and Ali e-commerce 29 marketplace for assessing public concerns, risk perception, and tracking public behaviors in response to the COVID-19 outbreak. More specifically, public emotion is evaluated by using a text analysis program, called Linguistic Inquiry and Word Count (LIWC), whereas public attention and awareness, and misinformation are assessed by constructing a Weibo daily index, which is the number of posts with keywords related to the COVID-19 pandemic. Moreover, the Baidu and Ali daily indices are used to evaluate the intention and behaviors to follow the recommended protection measures and/or rumors about ineffective treatments during the COVID-19 outbreak.
The results show that quickly classifying rumors and misinformation can greatly alleviate the impact of irrational behavior. Applications computer audition (CA), i.e., speech and sound analysis with AI, to contribute to the COVID-19 pandemic, were considered in \cite{schuller2020covid}. Similar to textual cues in \cite{hou2020assessment}, it is possible to collect spoken conversations from news, advertising videos, and social medias for further analyses. Several potential use-cases are also presented, including risk assessment, diagnosis, monitoring the spread and social distancing effects, checking the treatment and recovery, generation of speech and sound. Along the potentials of CA  in the fight against COVID-19, there are some challenges needed to be addressed, for example, how to collected COVID-19 patient data, how to process the audio and speech data effectively and in real-time, and how to explain the results obtained from the CA-based solutions. 

Another application of AI can be found in \cite{ye2020alpha}, where the COVID-19 related data is collected from heterogeneous sources at multiple levels, including official public health organizations (e.g., WHO and county government websites), demographic data, mobility data (e.g., traffic density from Google map), and user generated data from social media platforms. By utilizing the conditional generative adversarial nets to enrich the limited data and a novel heterogeneous graph auto-encoder to estimate the risk in an hierarchical
fashion. The proposed $ \alpha $-Satellite system allows us to be aware of the COVID-19 risk in a specific location, thus enabling the selection and appropriate actions to minimize the impact of COVID-19.
Recently, the authors in \cite{liu2020machine} proposed a novel method, namely Augmented ARGONet, to estimate the number confirmed COVID-19 cases in two days ahead. The data is collected from multiple sources, including official health reports from China CDC, COVID-19 Internet search activities from Baidu, news media activities from Media Cloud, and  daily forecasts achieved by the model proposed in \cite{Chinazzieaba9757}. To enrich the dataset, each data point is also augmented by adding a random Gaussian noise with mean $ 0 $ and standard deviation $ 1 $. The Lasso regression model is then applied to predict the number of confirmed cases for 32 provinces in China. The results show that the proposed Augmented ARGONet is able to outperform the baseline models for most testing scenarios.

\subsection{AI for biomedicine and pharmacotherapy}
The world has seen a race for finding effective vaccines and medical treatments in order to combat the COVID-19 virus, which demands for substantial efforts from not only health science but also computer science, with the help of AI and emerging technologies. 
The authors in \cite{mamoshina2016applications} have answered the question ``\textit{Why AI may benefit biomedical research}". First, the vast amount of biomedical data has triggered the use of AI in various areas of biomedicine and pharmaceutical industry. DL with deep neural networks is able to handle high-dimensional and non-structured data with nonlinear relationships, which are in line with biomedical data like transcriptomics and proteomics. Lastly, high capabilities in extracting higher-level features makes DL a potential candidate to analyze, bind, and interpret heterogeneous medical data such as DNA microarray data. The past few years have seen a wide range of AI applications to biomedicine researches, for example, medical image classification, genomic sequence analysis, classification and prediction, biomarkers, structural biology and chemistry, multiplatform data processing, drug discovery and repurposing 
\cite{Cao201817Deep, ekins2019exploiting}. 
Due to the severity of COVID-19 pandemic, AI has recently found medicine applications to curb the spread of the coronavirus, which mainly focus on protein structure prediction, drug discovery, and drug repositioning.

\begin{table*}[t]
	\color{black}{
	\caption{\textcolor{black}{Summary of the state-of-the-art studies on AI applications for COVID-19.}}
	\label{Table:Summary_AI_Applications}
	\resizebox{\textwidth}{!}{	
		\begin{tabular}{|c|c|p{13.5cm}|}
			\hline 
			Category & Paper & Highlights \& Contributions  \\ 
			\hline
			
			\multirow{12}{*}{\makecell{Detection \\and \\diagnosis}} & \multirow{4}{*}{\cite{abbas2020classification}} & A CNN-based DeTraC framework is proposed. In particular, the transfer learning concept is used to utilize well-performed deep models. \\
			& & For the pre-trained ResNet18 model, the DeTraC method achieves competitive performance, accuracy of 95.12\%, sensitivity of 97.91\%, and specificity of 91.87\%.\\
			
			\cline{2-3} 
			
			&  \multirow{2}{*}{\cite{wang2020covid}} & A deep CNN model for classification of COVID-19 and the dataset is designed by collecting $ 13,975 $ chest X-ray images across $ 13,870 $ patients. The proposed CNN model can achieve the test accuracy of 93.3\%.\\ 
			
			\cline{2-3} 
			
			& \multirow{2}{*}{\cite{tang2020severity}} &  Using chest CT images, 63 quantitative features of COVID-19 are analyzed by an RF model.\\ 
			&& The proposed method can obtain promising results, e.g., accuracy of 0.875 and AUC score of 0.91.\\ 
			
			\cline{2-3} 
			
			& \multirow{4}{*}{\cite{imran2020ai4covid19}} &  The AI4COVID-19 framework is proposed to consider domain knowledge of medical experts. The input data is cough/sound signals, which may be recorded by smartphones. \\
			&& The performance is very promising, the classification accuracy of 97.91\% (93.56\%) is obtained for cough (COVID-19) detection. \\ 
			
			\hline
			
			\multirow{5}{*}{\makecell{Identifying, \\tracking, \\and predicting \\the outbreak}} & \multirow{2}{*}{\cite{chen2020time}} & A time-dependent SIR model is proposed to dynamically adjust the control parameters according to the outbreak policies. The model is also extended to consider undetectable infected cases.\\
			
			\cline{2-3} 
			
			& \multirow{3}{*}{\cite{hu2020forecasting}} & A modified autoencoder framework is investigated to model the transmission dynamics of COVID-19. Using the empirical data from the WHO, the model can achieve an average error of less than 2.5\%. \\
			&& An interesting observation is that a faster intervention can significantly reduce the numbers of infected and dead cases. \\ 
			
			\hline
			
			\multirow{8}{*}{\makecell{Infodemiology \\and \\infoveillance}} & \multirow{4}{*}{\cite{hou2020assessment}} & Data is collected from  Sina Weibo, Baidu search engine, and Ali e-commerce 29 marketplace to evaluate public concerns/behaviors and risk perception to the COVID-19 outbreak. \\
			&& The result shows that fast reaction to quickly classify rumors and misinformation can well mitigate impacts of irrational behaviors.\\
			
			\cline{2-3} 
			
			& \multirow{2}{*}{\cite{schuller2020covid}} & Applications of computer visions for combating the COVID-19 pandemic are presented. Potential use cases (e.g., risk assessment and diagnosis) and challenges (e.g., data collection and model sharing) are also discussed.\\ 
			
			\cline{2-3}
			
			& \multirow{2}{*}{\cite{ye2020alpha}} & An AI-driven system, namely $ \alpha $-satellite, is proposed to estimate the risk of COVID-19 in an hierarchical manner. Data is collected from heterogeneous sources, e.g., WHO, demographic and mobility data, and social platforms.\\ 
			
			\hline
			
			\multirow{6}{*}{\makecell{Biomedicine \\and \\pharmacotherapy}} & \multirow{2}{*}{\cite{hu2020prediction}} & A pre-trained deep learning model is utilized to train a dataset of $ 4,895 $ commercially available drugs.\\ 
			&& After learning and manual refinement, 10 drugs are selected as potential COVID-19 inhibitors. \\
			
			\cline{2-3} 
			
			& \multirow{2}{*}{\cite{ge2020data}} & For drug repurposing, a data-driven approach is examined over $ 6,000 $ candidate drugs. The key finding is that the inhibitor CVL218 is very promising and has a safety profile in monkeys and rats.\\ 
			
			\cline{2-3}
			
			& \multirow{3}{*}{\cite{chenthamarakshan2020targetspecific}} & A deep generative model, namely CogMol, is proposed to find potential molecules that can blind three relevant protein structures of coronavirus. \\
			&& Also, \textit{in silico} screening experiments are conducted to assess the toxicity of the generated molecules.\\ 
			
			\hline
		\end{tabular}
	}
}
\end{table*}

Using AI to discover new medicines and/or existing drugs to be immediately used to treat the COVID-19 virus is beneficial from both the economic and scientific perspectives, especially when clinically verified medicines and vaccines are not available. 
The work in \cite{hu2020prediction} leveraged a COVID-19 virus-specific dataset to train a DL model. Then, this trained model is used to check $ 4,895 $ commercially available drugs and find potential inhibitors with high affinities. Overall, ten existing drugs are listed as potential inhibitors, for example, the HIV inhibitor \textit{abacavir}, and respiratory stimulants \textit{almitrine mesylate and roflumilast}.
The work in \cite{ge2020data} proposed a data-driven model for drug repurposing through the combination of ML and statistical analysis methods. Initially, the authors selected a list of $ 6225 $ candidate drugs, which are then narrowed through consecutive phases, including a network-based knowledge mining algorithm, a DL based relation extraction method, and a connectivity map analysis approach. After the \textit{in silico} and \textit{in vitro} studies, a novel1 inhibitor, namely  CVL218, has been found to be a drug candidate to treat the COVID-19 virus. The potential of this finding is that the inhibitor CVL218 is verified to have a safety profile in monkeys and rats. Another interesting work for drug repurposing was in \cite{savioli2020oneshot}, which leverages a Siamese neural network (SNN) to identify the COVID-19 protein structure versus Ebola and HIV-1 viruses. Advantages this work is that the proposed DL model can be trained without the need for a large dataset and can work directly with available biological dataset instead of public datasets, which are not specific about the COVID-19 virus. Besides drug repurposing, many have been devoted to drug discovery \cite{Ton2020Rapid, chenthamarakshan2020targetspecific, hofmarcher2020large, Ong2020COVID19} and protein structure prediction \cite{Jumper2020Computational, senior2020improved, Strokach868935} for combating against the COVID-19 pandemic. For example, the work in \cite{chenthamarakshan2020targetspecific} considered a DL model named controlled generation of molecules (CogMol) to learn candidate molecules that can bind protein targets of the COVID-19 virus, which are then used to generate candidate drugs for the treatment of the COVID-19 virus. Moreover, a multitask deep neural network is used to predict the toxicity of the generated molecules, thus improving the \textit{in silico} screening process and increasing the success rate of the drug candidates.

\subsection{Summary of AI Applications for COVID-19}
In this section, we review the applications of AI for {COVID-19} detection and diagnosis, tracking and identification of the outbreak, infodemiology and infoveillance, biomedicine and pharmacotherapy. We observe that the AI-based framework is highly suitable for mitigating the impact of the COVID-19 pandemic as a vast amount of COVID-19 data is becoming available thanks to various technologies and efforts. Through the AI studies are not implemented at a large scale and/or tested clinically, they are still helpful as they can provide fast response and hidden meaningful information to medical staffs and policy makers. However,
we face many challenges in designing AI algorithms as the quality and quantity of COVID-19 datasets should be further improved, which call for constant effort from the research communities and the help from official organizations with more reliable and high-quality data. A summary of state-of-the-art studies on AI techniques for is summarized in Table~\ref{Table:Summary_AI_Applications}. 
\section{Applications of Big Data for Fighting COVID-19}
\label{Sec:Applications_BD}
\begin{figure*}[t]
	\centering
	\includegraphics[width=0.825\linewidth]{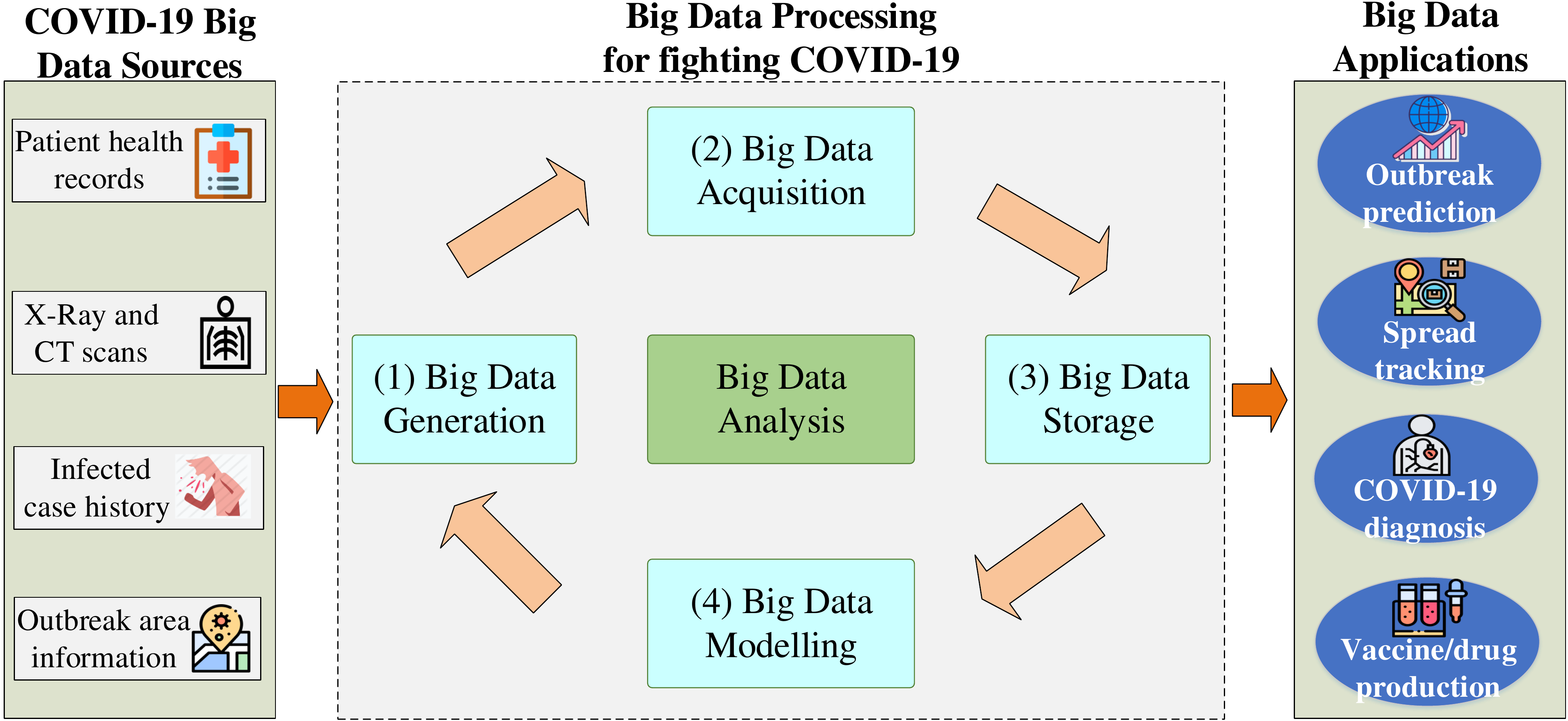}
	\caption{Big data and its applications for fighting COVID-19 pandemic.}
	\label{Fig:Bigdata_COVID_app}
\end{figure*}

In the sense of COVID-19 data storage and analysis, big data analytics for COVID-19 is mainly characterized via some key techniques from the big data concept, such as multi-domain dataset analysis (i.e. combined COVID-19 image-time series data), high-dimensional analysis (i.e. 3-D COVID-19 images), deep analysis (i.e. using deep Boltzmann machine, hidden Markov model, deep neural network), and parallel computing (i.e. bit-, instruction-, data-, and task-parallelism). In this section, we summarize some key big data use cases and applications enabled by these techniques, including outbreak prediction, virus spread tracking, coronavirus diagnosis/treatment, and vaccine/drug discovery that are summarized in Fig.~\ref{Fig:Bigdata_COVID_app}. 


\subsection{Outbreak prediction}
Big data plays an important role in combating COVID-19 through its capability to predict the outbreak from large-scale data analytics. For example, the work in \cite{20} leveraged real datasets collected from the COVID-19 pandemic in Italy to estimate the outbreak possibility which is of significant importance to plan effective disease control strategies. Instead of using a simple and deterministic model based on human transmission \cite{21}, the authors developed more complex models that can accurately formulate the dynamics of the pandemic based on huge data sets from Italian Civil Protection sources. Another data source for outbreak prediction is the public dataset that can be used to visualize the geographical areas with possible outbreak \cite{22}. A first trial is an investigation in Wuhan, aiming to monitor the people migration from and to Wuhan city so that the health agencies can predict the population infected with COVID-19 for quarantine. 

Meanwhile, the work in \cite{23} employed the COVID-19 outbreak-related data from authoritative sources such as the National, provincial and municipal Health Commissions of China (\url{http://www.nhc.gov.cn/}). This big data source can help implement pandemic modeling to interpret the cumulative numbers of infected people, recovered cases in five different regions, i.e., the Mainland, Hubei, Wuhan, Beijing, and Shanghai. More importantly, it allows us to perform simulations to predict the tendency of the COVID-19 outbreak, i.e. identifying the areas at high risks of pandemic and detecting the population with increasing infected cases, all of which contribute to the success of anti-pandemic campaigns. 

Big data also enables outbreak prediction on the global scope. From the data analysis point of view, the outbreak is predicted using available data points that put the accuracy of the fitting for reliable estimation under doubt due to the lack of comprehensive investigations. Accuracy may depend on the number of contributed factors, ranging from infected cases, population, living conditions, environments, etc. Motivated by this, a research effort in \cite{24} leveraged a large dataset from various regions and countries such as Korea, China, to estimate the pandemic based on a logistic model that can adjudge the reliability of the predictions. Another trial in \cite{25} used Google Trends as a data engineering tool to collect coronavirus-related information in China, South Korea, Italy, and Iran. The data comes from a textual search with five geographical settings: 1) worldwide, to investigate the global interest in coronaviruses; 2) China, where there is currently the highest number of  infected cases; 3) South Korea, where interest has increased since 19 February because hundreds of new infected cases were confirmed; 4) Italy; and 5) Iran, where since 22 February hundreds of new cases have been detected. This data source aggregation would help visualize the outbreak tendency and estimate the possible outbreak. The government reports collected from the coronavirus pandemics in China, South Korea, Italy, and Iran are also collected and analyzed by \cite{26} with a data optimization model, aiming to generate an accurate prediction of the daily infected cases, which potentially long-term predictions on the future outbreak. 

Moreover, the work in \cite{27} employed datasets from John Hopkins University repository, which are obtained from confirmed cases, death cases and recovered cases of all countries, to build prediction models using data learning. The trial is implemented in India, indicating that the proposed data analytic can estimate the outbreak in short-term intervals, i.e. two-week duration, which potentially expands to  build larger models for long-term predictions. Meanwhile, another data analytic method in \cite{28} was investigated in the US with the large-scale datasets collected from American cities. The key purpose of the study is learning from recent data to calculate prediction errors to optimize the data modeling model for improving the quality of future estimation, possibly for coronavirus-like pandemics. 

\subsection{Virus spread tracking}
Another role of big data is the tracking of COVID-19 spread, which is of paramount importance for healthcare organizations and governments in controlling successfully the coronavirus pandemic \cite{29}. A number of newly emerging solutions using big data have been proposed to support tracking of the COVID-19 spread. For instance, the study in \cite{30} suggested a big data-based methodology for tracking the COVID-19 spread. A large dataset is collected from China National Health Commission with 854,424 air passengers left 55 Wuhan Airport to 49 cities in China during the period of December 2019-January 2020. A multiple linear model is built using local population and air passengers as estimated variables that are useful to quantify the variance of reported cases in China cities.  More specifically, the authors employed a Spearman correlation analysis for the daily user traffic from Wuhan and the total user traffic in this period with the number of 49 confirmed cases. The analytic results show a high correlation between the positive infection cases and the population size. The authors in \cite{31} considered leveraging both big data for spatial analysis methods and Geographic Information Systems (GIS) technology which would facilitate data acquisition and the integration of heterogeneous data from health data resources such as governments, patients, clinical labs and the public.

Another research effort in \cite{32} combined datasets collected from China, Singapore, South Korea, and Italy to build a comprehensive analytic model for virus spread tracking. Based on data learning and modelling, a macroscopic growth law is derived that allows us to estimate the maximum number of infected patients in a certain area. This is important for effective assessment of COVID-19 prevention and monitoring the potential spread of COVID-19 disease, especially the nearby regions of the central epidemic. Different from \cite{32}, the study in \cite{33} proposed a temperature-based model that evaluates the relationship between the number of infected cases and the average temperature in different countries necessary for coronavirus tracking. A large dataset is collected from 42 countries which developed the epidemic earlier and built a dataset of 88 countries. The analysis results from COVID-19 tracking show that in northern hemisphere countries, the growth rate should significantly decrease due to the warmer weather and lockdown policies, compared to the southern hemisphere countries.

The previous works have focused on big data solutions, where ground truth data, in the form of historical syndromic surveillance reports, can be utilized to build data models for disease tracking. However, the accuracy and reliability of the modelling models are under questions. The authors in \cite{34} suggested an unsupervised model for COVID-19 spread tracking from online data. They selected a wide range of related symptoms as identified and collected by a survey from the National Health Service (NHS) in the United Kingdom (UK) by incorporating a basic news media coverage metric associated with confirmed COVID-19 cases. Then, a transfer learning method is proposed for mapping supervised COVID-19 models from a country to another country where the COVID-19 epidemic possibly spreads.

\subsection{Coronavirus diagnosis/treatment}
In addition to outbreak prediction and spread tracking applications, big data has the potential to support COVID-19 diagnosis and treatment processes. In fact, the potential of big data for diagnosing infectious diseases like COVID-19 has been proved via recent successes, from early diagnosis \cite{37}, prediction of treatment outcomes, and building supportive tools for surgery. Regarding COVID-19 diagnosis and treatment, big data has provided various solutions as reported in the literature. A study in \cite{40} introduced a robust, sensitive, specific and highly quantitative solution based on multiplex polymerase chain reactions that are able to diagnose the SARS-CoV-2. The model consists of comprised of 172 pairs of specific primers associated with the genome of SARS-CoV-2 that can be collected from the Chinese National Center for Biological information (https://bigd.big.ac.cn/ncov). The proposed Multiplex PCR scheme has been shown to be an efficient and low-cost method to diagnose Plasmodium falciparum infections, with high coverage (median 99\%) and specificity (99.8\%). Another research in \cite{41} implemented a molecular diagnostic method for genomic analyses of SARS-CoV-2 strains, with a focus on studying on Australian returned travelers with COVID-19 disease using genome data available at \url{https://www.gisaid.org/}. This study may provide important insights into viral diversity and support COVID-19 diagnosis in areas with lacking genomic data. 

Meanwhile, the authors in \cite{42} used the proteomics cells that get infected with COVID-19 virus for diagnosis. More specifically, data includes 6381 proteins in human cells with the COVID-19 infection. Based on that, a cooperative analysis of an impact pathway analysis and a network analysis has been derived to analyze the data gathered from the Kyoto Genes storage. Another solution in \cite{43} leveraged publicly available mouse and nonhuman primate for non-neural expression of SARS-CoV-2 entry genes. As a trial, a bulk and single-cell RNA-Seq dataset is tested to detect all types of cells that got infected with COVID-19 virus, which is necessary for diagnosis and/or prognosis in COVID-19. 

Due to the limited dataset for COVID-19 diagnosis and treatment, more experimental data has been generated in \cite{44} where all the 6 epitopes (A, B, C, E, F/G and H) can induce the body to produce corresponding antibodies and generate specific humoral immunity. The sequence of S, E, M protein and its proximal sequences have been employed, which produced 420, 334 and 329 sequences in total from NCBI database to build the final dataset after genome annotation. Based on that, the authors focus on a diagnosis procedure by the prediction of the 3D structure of target protein and prediction of conformational B cell epitopes of target protein of SARS-CoV-2 coupled with an analysis of epitope conservation. 

Interestingly, the recent work in \cite{45} presented a comprehensive guideline with useful tools to serve the diagnosis and treatment of COVID-19. This guideline consists of the guideline methodology, epidemiological characteristics, population prevention, diagnosis, treatment of COVID-19 disease. As a first trial, a large-scale data of Zhongnan Hospital of Wuhan University has been analyzed from a collection system where 11,500 persons were screened, and 276 were identified as suspected infectious victims, and 170 were diagnosed. An array of clinical tests have been implemented from the big dataset, from Typical and Atypical CT/X-ray imaging manifestation to hematology examination and detection of pathogens in the respiratory tract. 

\subsection{Vaccine/drug discovery }
Developing a novel vaccine is very crucial to defending the rapid endless of global burden of the COVID-19 pandemic. Big data can gain insights for vaccine/drug discovery against the COVID-19 pandemic. Few attempts have been made to develop a suitable vaccine for COVID-19 using big data within this short period of time. The work in \cite{46} used the GISAID database (www.gisaid.org/CoV2020/) that has been used to extract the amino acid residues. The research is expected to seek potent targets for developing future vaccines against the COVID-19 pandemic.  Another effort for vaccine development is in \cite{47} that focused on  investigating the spike proteins of SARS CoV, MERS CoV and SARS-CoV-2 and four other earlier out-breaking human coronavirus strains. This analysis would enable critical screening of the spike sequence and structure from SARS CoV-2 that may help to develop a suitable vaccine.

The approaches of reverse vaccinology and immune informatics can be useful to develop subunit vaccines towards COVID-19. \cite{48}. The strain of the SARS-CoV-2 has been selected by reviewing numerous entries of the online database of the National Center for Biotechnology Information, while the online epitope prediction server Immune Epitope Database was utilized for T-cell and B-cell epitope prediction. An array of steps may be needed to analyze from the built dataset, including antigenicity, allergenicity and physicochemical property analysis, aiming to identify the possible vaccine constructs. A huge dataset has been also collected from the National Center of Biotechnology Information for facilitating vaccine production \cite{49}. Different peptides were proposed for developing a new vaccine against COVID-19 via two steps. First, the whole genome of COVID-19 was analyzed by a comparative genomic approach to identify the potential antigenic target. Then, an Artemis Comparative Tool has been used to analyze human coronavirus reference sequence that allows us to encode the four major structural proteins in coronavirus including envelope (E) protein, nucleocapsid (N) protein, membrane (M) protein, and spike (S).

Big data also helps provide strategies for drug manufacturing for fighting COVID-19. For example, a solution based on molecular docking was proposed in \cite{50} for drug investigations. More than 2500 small molecules in the FDA-approved drug database were first screened and validated through a molecular docking program called Glide. As a result, fifteen out of twenty-five drugs validated in exhibited significant inhibitory potencies, inhibiting signaling pathway and inflammatory responses and prompting drug repositioning against COVID-19. A big data-driven drug repositioning scheme was also introduced in \cite{51}. The key purpose of this project is to apply ML to combine both knowledge graph and literature for COVID-19 vaccine development. 

\begin{table*}[t]
		\color{black}{
		\caption{\textcolor{black}{Summary of the state-of-the-art studies on big data applications for COVID-19.}}
		\label{Table:Summary_BD_Applications}	
			\resizebox{\textwidth}{!}{	
			\begin{tabular}{|c|c|p{13.5cm}|}
				\hline 
				Category & Paper & Highlights \& Contributions  \\ 
				\hline
				
				\multirow{12}{*}{\makecell{Outbreak prediction}} & \multirow{4}{*}{\cite{20}} & A big data platform is proposed to estimate the outbreak possibility using the huge data sets from Italian Civil Protection sources. \\ && The first trial is implemented in Wuhan to predict the population infected with COVID-19 for quarantine. \\
				
				\cline{2-3} 
				
				&  \multirow{2}{*}{\cite{23}} & A big data-based solution is proposed to implement pandemic modeling to interpret the cumulative numbers of infected people, recovered cases in different regions, i.e., Wuhan, Beijing, and Shanghai. \\ && This scheme is able to predict the tendency of the COVID-19 outbreak in the areas at high risks of pandemic. \\ 
				
				\cline{2-3} 
				
				& \multirow{2}{*}{\cite{24}} &  A framework is introduced using a large dataset from various regions and countries such as Korea, China, to estimate the pandemic based on a logistic model that can adjudge the reliability of the predictions. \\ 
				
				\cline{2-3} 
				
				& \multirow{4}{*}{\cite{28}} &  A big data analytic method is investigated in the US with the large-scale datasets collected from American cities. \\ && The approach enables to calculate prediction errors to optimize the data modeling model for improving estimation accuracy. \\
		 		\hline
		 		
				\multirow{12}{*}{\makecell{Virus spread tracking}} & \multirow{4}{*}{\cite{30}} & A big data-based analytic methodology for tracking the COVID-19 spread is considered using a large dataset collected from China National Health Commission with 854,424 people. \\ && The analytic results show a high correlation between the positive infection cases and the population size.\\
				
				\cline{2-3} 
				
				&  \multirow{2}{*}{\cite{32}} & A big data-based analytic model is built using datasets collected from China, Singapore, South Korea, and Italy for virus spread tracking.  \\ && This model can estimate the maximum number of infected patients in a certain area.  \\ 
				
				\cline{2-3} 
				
				& \multirow{2}{*}{\cite{33}} & A temperature-based model is proposed to evaluate the relationship between the number of infected cases and the average temperature in different countries necessary for coronavirus tracking. \\ 
				
				\cline{2-3} 
				
				& \multirow{4}{*}{\cite{34}} &  A big data-based unsupervised model is designed for COVID-19 spread tracking from online data by incorporating a basic news media coverage metric associated with confirmed COVID-19 cases. \\ && The work is in progress for coronavirus tracking tasks.  \\
				\hline
				\multirow{10}{*}{\makecell{Coronavirus diagnosis/ \\treatment}} & \multirow{4}{*}{\cite{40}} & A robust, sensitive, specific and highly quantitative solution based on multiplex polymerase chain reactions is proposed to diagnose the SARS-CoV-2. \\ && The proposed scheme has been shown to be an efficient and low-cost method to diagnose Plasmodium falciparum infections.  \\
				
				\cline{2-3} 
				
				&  \multirow{2}{*}{\cite{42}} & A method is proposed using 6381 proteins in human cells that get infected with COVID-19 virus. \\ && This aims to analyze the data gathered from the Kyoto Genes storage to serve COVID-19 diagnosis. \\ 
				
				\cline{2-3} 
				
				& \multirow{2}{*}{\cite{45}} &  An array of clinical tests have been implemented from the big dataset, from Typical and Atypical CT/X-ray imaging manifestation to hematology examination and detection of pathogens in the respiratory tract. \\ && These tests provide a comprehensive guideline with useful tools to serve the diagnosis and treatment of COVID-19. 
				\\ 
				\hline
				\multirow{7}{*}{\makecell{Vaccine/drug discovery}} & \multirow{4}{*}{\cite{47}} & A method is proposed to investigate the spike proteins of SARS CoV, MERS CoV and SARS-CoV- 2 and four other earlier out-breaking human coronavirus strains. \\ && It enables critical screening of the spike sequence and structure from SARS CoV-2 for vaccine development.  \\
				
				\cline{2-3} 
				
				&  \multirow{2}{*}{\cite{49}} & A project is built using a huge dataset collected from the National Center of Biotechnology Information for facilitating vaccine production. Different peptides were proposed for developing a new vaccine against COVID-19. \\ 
				
				\cline{2-3} 
				
				& \multirow{2}{*}{\cite{50}} &  A solution is proposed based on molecular docking for drug investigations with over 2500 small molecules, which aims prompting drug repositioning against COVID-19. \\ 
		
				\hline
			\end{tabular}}}
\end{table*}

\subsection{Summary of Big Data Applications for COVID-19}
Reviewing from the rapidly emerging literature, we find that big data plays an important role in combating the COVID-19 pandemic through a number of promising applications, including outbreak prediction, virus spread tracking, coronavirus diagnosis/treatment, and vaccine/drug discovery. In fact, big data potentially enables outbreak prediction on the global scope using data analytic tools on huge datasets collected from available sources such as health organizations (e.g. WHO), healthcare institutes (i.e. China National Health Commission) \cite{24}, \cite{25}. Big data has also emerged as a promising solution for coronavirus spread tracking by combining with intelligent tools such as ML and DL \cite{32} for building prediction models that is very useful for governments in monitoring the potential COVID-19 outbreak in the future. Moreover, big data has the potential to support COVID-19 diagnosis and treatment processes. The research results from the literature studies prove that big data can help healthcare provide  various medical operations from early diagnosis, disease analysis, and prediction of treatment outcomes \cite{40}, \cite{41}. Finally, data learning from big datasets also helps finding potential targets for an effective vaccine against SARS-CoV-2 \cite{47}, and integrating large-scale knowledge graph, literature and transcriptome data, supporting to discover the potential drug candidates against SARS-CoV-2 \cite{51}. A summary of big data applications for COVID-19 is shown in Table~\ref{Table:Summary_BD_Applications}. 
\section{Examples of AI and Big Data Frameworks for Covid-19}
\label{Sec:Examples}

\begin{figure}[t]
	\centering
	\includegraphics[width=0.95\linewidth]{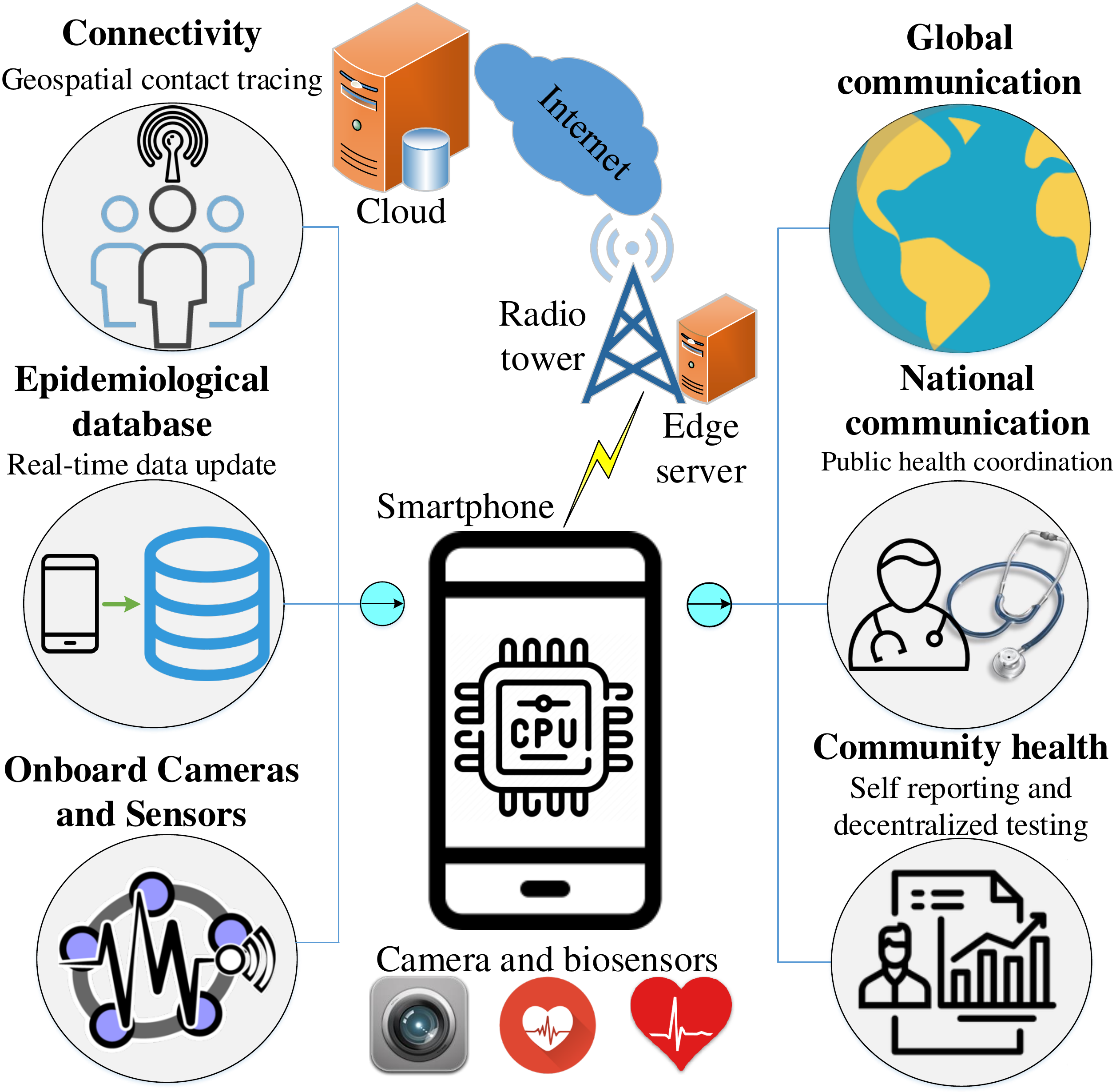}
	\caption{An AI-based framework using mobile phones for COVID-19 diagnosis and surveillance. Adapted from \cite{udugama2020diagnosing}.}
	\label{Fig:AI_Based_Phone_FR}
\end{figure}

This section presents two examples of AI- and big data based frameworks of how AI and big data are currently helping in the fight against the COVID-19 pandemic

\subsection{Phone-based solution for detection and surveillance}

Thanks to recent advancements in hardware capabilities, evolution of wireless and mobile communications, and emergence of edge/cloud computing, mobile phone have many opportunities during the outbreak, e.g., outbreak identification, detection and diagnosis, treatment and infected case management, and disease elimination. Adapted from \cite{udugama2020diagnosing}, we propose a mobile phone based framework for COVID-19 detection and surveillance, as shown in Fig.~\ref{Fig:AI_Based_Phone_FR}. Wireless connectivity, e.g., WiFi, 3G, 4G, and 5G, has been widely deployed in almost all the countries, so every individual with a mobile phone can connect to the Internet and can do many activities online. The DL model can be trained in the cloud, even at a server collocated at the network edge, which is then pushed to mobile phone for further purpose. For example, by using the embedded cameras and biosensors, a mobile phone can collect personal information, for example, X-ray and CT images, cough sound, and heart rate, which may be encrypted and compressed before sending to the cloud for training \cite{imran2020ai4covid19, rao_vazquez_2020}. Review on recent advances in biosensors for mobile health platforms can be found in 
\cite{geng2017recent}. 
The massive number of mobile devices connected to the Internet may relax the limited data sent from an individual phone. According to the latest annual Internet report from Cisco \cite{CiscoReport}, by 2023 around 29.3 billion networked devices will be connected to the Internet, where mobile phones and tablets account for $ 23 $\% and $ 3 $\% respectively, and over 70\% of the world population will have mobile connectivity. 

Despite many potentials, a number of challenges need to be addressed for the success of AI-based mHealth \cite{wood2019taking}. The very first challenge is from the capability and reliability of hardware and software, which are used in mobile devices to the medical purpose. This challenge can be overcome by developing standard mobile devices with specific computing/sensing components and system on chip (SoC). Another challenge arises from the protocol to collect/store the diagnosis results, and the path way to improve the patients' trust, especially when they normally prefer to have a face-to-face talk with doctors and clinicians. For example, depending on local and national policies, the diagnostic results collected from individuals in a region are required to transmit to and store at the local server instead of a cloud server, which would be used to train the DL model for global communication. Lastly, a challenge in utilizing mobile devices for the detection and surveillance is how to evaluate the clinical cost and effectiveness. This is reasonable since the results using phone-based frameworks are usually questionable and not comparable with those of the direct tests.

\subsection{AI and big data for neutralizing antibody discovery}

\begin{figure}[t]
	\centering
	\includegraphics[width=0.95\linewidth]{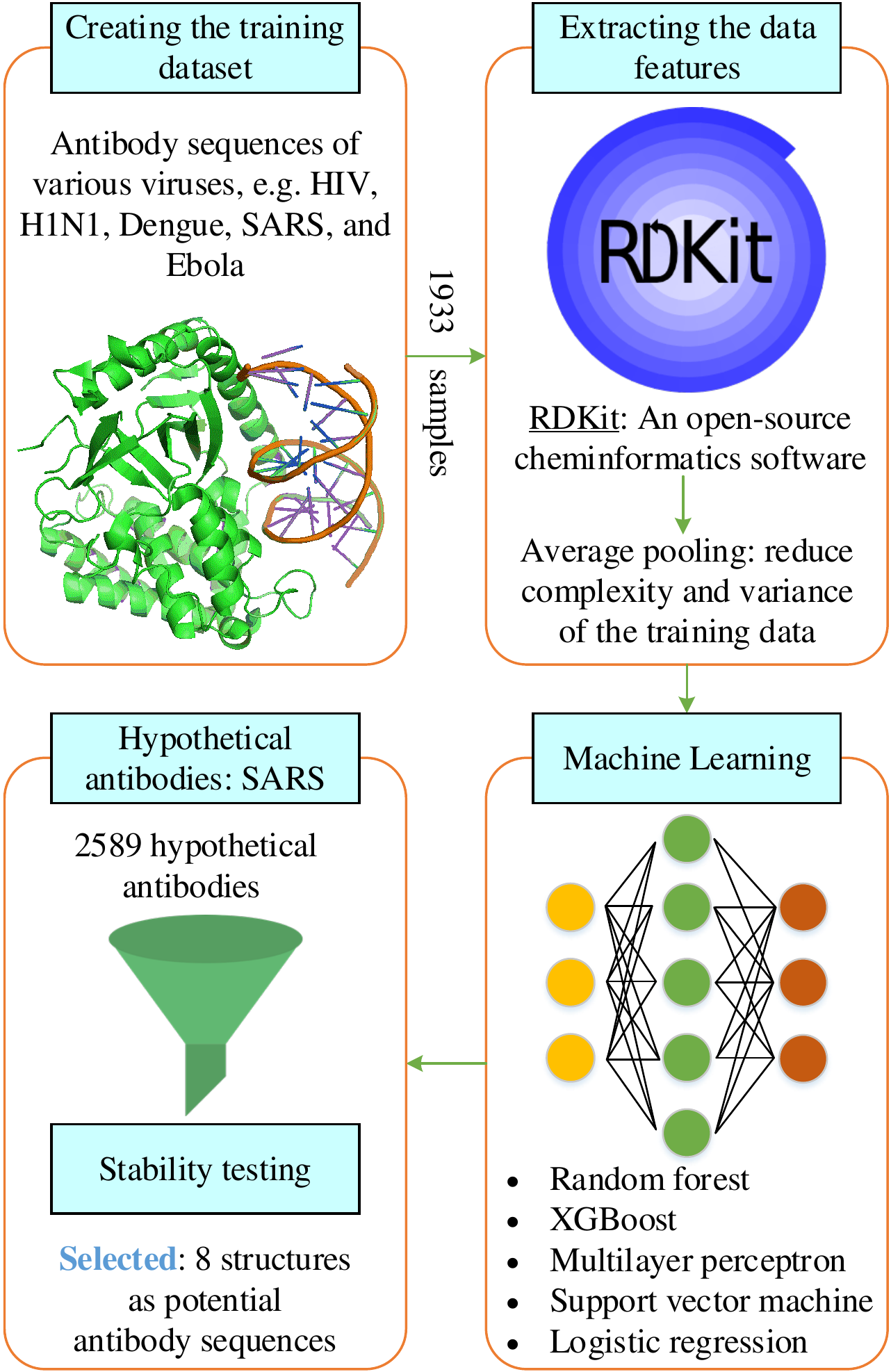}
	\caption{Illustration of a data-driven framework for discovering antibody sequences to treat the COVID-19 disease \cite{magar2020potential}.}
	\label{Fig:AI_Bigdata_AntibodySequences}
\end{figure}

Under the current situation that specific medicines and vaccines for COVID-19 virus are not available, it is urgent to find efficient and fast solution to combat the COVID-19 disease and prevent the virus outbreak. Motivated by this, the authors in \cite{magar2020potential} introduced a data-driven framework that combines AI, big data, and medical knowledge to identify antibody sequences that can inhibit the growth of COVID-19 virus. The schematic illustration of the framework proposed in \cite{magar2020potential} is pictorially shown in Fig.~\ref{Fig:AI_Bigdata_AntibodySequences}. The initial dataset is consisted of 1831 antigen and antibody sequences of various viruses, for example, HIV, H1N1, Dengue, SARS, and Ebola from the CATNAP tool \cite{yoon2015catnap}, and their corresponding half maximal inhibitory concentration (IC50) values\footnote{IC50 is a quantitative measure that denotes the concentration of a substance required for $ 50 $\% inhibition \cite{ref1}.}. Moreover, the authors proposed mining more $ 102 $ data samples of from the RCSP protein data bank \cite{berman2003protein} (URL: \url{https://www.rcsb.org/}), which is then added to the initial data set, thus creating the final dataset of $ 1933 $ samples. 

The next phase is extracting the features from the dataset of $ 1933 $ samples. In doing so, the authors exploited an open-source cheminformatics software (available at \url{https://www.rdkit.org/}), namely RDKit, to represent the molecular graph. Moreover, to reduce computational complexity and variance of the training data and, the authors applied an average-pool layer over the extracted features. Five ML models, including random forest, XGBoost, multilayer perceptron, support vector machines, and logistic regression, are then applied to evaluate their performance in terms of five-fold cross validation. The experiment results show that the XGBoost model is the most superior one, which is therefore selected to find the potential antibody candidates for COVID-19 virus. More importantly, the XGBoost model is highly promising as it can achieve an outstanding out-of-class prediction over various virus families, for example, $ 100 $\% for SARS and Dengue, $ 84.61 $\% for Influenza, and $ 75 $\% Ebola and Hepatitis.
After that, since COVID-19 virus has similar characteristics ($ 88 $-$ 89 $\% similarity \cite{LAI2020105924}) with SARS-CoV-2, the authors proposed generating a set of $ 2589 $ hypothetical candidates based on neutralizing antibodies and sequence features against SARS. Out of all the candidates, finally 8 structures are selected as potential antibody sequences for neutralizing the COVID-19 disease. The results discovered in this study would be used for further academic researches as well as find effective medicines and vaccines for the treatment of COVID-19. 
\section{Challenges, Lessons, and Recommendations}
\label{Sec:Lessons}
As reviewed in the previous sections, AI and big data have found their great potentials in the globe battle against the COVID-19 pandemic. Apart from the obvious advantages, there are still challenges needed to be discussed and addressed in the future. Additionally, we highlights some lessons stemmed from this paper and give some recommendations for the research communities and authorities. 

\subsection{Challenges and solutions}
\subsubsection{Regulation}
As the outbreak is booming and the daily number of confirmed cases (infected and dead) increases considerably, various approaches have been taken to the control this outbreak, e.g., lockdown, social distancing, screen and testing at a large scale. In this way, regulatory authorities occupy a crucial role in defining policies that can encourage the involvement of residents, scientists and researchers, industry, giant techs and large firms, as well as harmonizing the approaches executed by different entities to avoid any barriers and obstacles in the way of preventing the COVID-19 disease.

Regarding this challenge, many attempts have been made from the first confirmed COVID-19 to current situation. An example is from the quarantine policy in Korea, effective from 1st April 2020. More specifically, all passengers entering Korea are required to be quarantined for 14 days at registered addresses or designated facilities. Moreover, all passengers are required to take daily self-diagnosis twice a day and send the reports using self-diagnosis apps installed in their mobile phones. An ``AI monitoring calling system" has been implemented by the Seoul Metropolitan Government to automatically check the health conditions of the people who do not have any mobile phones and/or have not installed the self-diagnosis apps \cite{Seoul2020AI}. 
Another effort is from the collaboration between the Zhejiang Provincial Government and Alibaba DAMO Academy to develop an AI platform for automatic the COVID-19 testing and analysis \cite{HowChina2020AI}.  

\subsubsection{Lack of standard datasets}
In order to make AI and big data platforms and applications a trustful solution to fight the COVID-19 virus, a critical challenge arises from the lack of standard datasets. As surveyed in the previous sections, many AI algorithms and big data platforms have been proposed, but they are not tested using the same dataset. For example, the algorithms in \cite{wang2020deep} and \cite{ozkaya2020coronavirus} are verified to respectively achieve accuracy of $ 82.9 $\%/$ 98.27 $\%, specificity of $ 80.5 $\% and $ 97.60 $\%, sensitivity of $ 84 $\% and $ 98.93 $\%. However, we cannot decide which algorithm is better for the virus detection since two datasets with different numbers of samples are used. Furthermore, most datasets found in the literature have been made thanks to individual efforts, e.g., the authors collect some datasets available in the Internet, then unify them to create their own dataset and evaluate their proposed algorithms. 

To overcome this challenge, government, giant firms, and health organizations (e.g., WHO and CDC) play a key role as they can collaboratively work for high-quality and big datasets. A variety of data sources can be provided by these entities, e.g., X-ray and CT scans from the hospitals, satellite data, personal information and reports from self-diagnosis apps. For example, the CORD-19 dataset \cite{CORD-19} has been made and led by the Georgetown’s Center for Security and other partners like Allen Institute for AI, Chan Zuckerberg Initiative, Microsoft Research, and National Institutes of Health.
Alternatively, Alibaba DAMO Academy has worked with many Hospitals in China to build AI systems for detecting the COVID-19 infected cases, where Alibaba DAMO Academy is responsible for designing AI algorithms and hospitals are responsible for providing CT scans of more than $ 5,000 $ confirmed COVID-19 cases. As reported in \cite{HowDAMO2020AI}, this system has been used by more than 20 hospitals in China thanks to its remarkable performance: an accuracy of $ 96 $\% within $ 20 $ seconds only. As individual efforts, an initiative to manage public datasets of COVID-19 medical images is available at \cite{kalkreuth2020covid19} and a global collection of open-source projects on COVID-19 is available at \url{http://open-source-covid-19.weileizeng.com/}.

\subsubsection{Privacy and security challenges}
Right now, important things are keeping people healthy and soon controlling the situation; however, how personal data secure and private is still needed and should be investigated. A showcase of this challenge is the scandal of the Zoom videoconferencing app over its security and privacy issues\footnote{Zoom admits user data ‘mistakenly’ routed through China: https://www.ft.com/content/2fc518e0-26cd-4d5f-8419-fe71f5c55c98}. In the pandemic, authorities may request their people to share their personal information, for example, GPS location, CT scans, diagnosis reports, travel trajectory, and daily activities, which is needed to control the situation, make up-to-date policies, and decide immediate actions. Data is a must to guarantee the success of any AI and big data platforms; however, normally people do not want to share their personal information, if not officially requested. There is a trade-off: privacy/security and performance.

Many technologies are available to solve the privacy and security issues during the COVID-19 pandemic. Here, we consider some potential solutions below, which can serve as research directives in the future.
\begin{itemize}
	
	\item \textit{Blockchain}: Basically, blockchain is defined as  a decentralized, immutable and public database, where each transaction is verified by all the nodes in the network, which is enabled by consensus algorithms \cite{nguyen2019blockchain}. Blockchain has found its success for various healthcare applications \cite{kuo2017blockchain}, so it is possible to deploy blockchain based solutions to improve the user security and data privacy during the COVID-19 outbreak period. MiPasa is one of the projects that combines two emerging technologies from IBM: blockchain and cloud computing \cite{MiPasa}. The purpose of this project is to provide reliable, accessible, and high-quality data to the communities.
	
	\item \textit{Federated learning (FL)}: Traditionally, the data should be collected and stored centrally to training DL models. Federated learning offers a new solution in which a majority of personal data is not required to be transmitted to the central server \cite{yang2019federated}. Applying to the AI-based framework using mobile phones for diagnosing COVID-19 case, each phone can train its own DL model using local data. Mobile phones transmit their trained models to a central server, which is responsible for aggregating to make a global model that is then disseminated to all the mobile phones. We note that a phone is not limited to mobile phone only, it can be a server at the local health department, whereas the aggregation server can be a global cloud like Microsoft Azure and Amazon Web Services. 
	
	\item \textit{Incentive mechanisms}: A large and reliable dataset lays a foundation for AI and big data platforms contend with the COVID-19 outbreak. Therefore, there is a need for incentive design to call the participation of more people and entities in contributing their own data. Incentives are needed due to the following reasons: 1) a massive quantity of data are available from people/entities, who may be not requested by the governments to provide their data, and 2) the quality of data should be guaranteed in order to improve the accuracy and performance of learning models. Such incentive mechanisms for healthcare, wireless communications, transportation, etc. can be found in \cite{gao2015survey}.
\end{itemize}


\subsection{Lessons and Recommendations}
Reviewing the state-of-the-art literature, we find that AI and big data technologies play a key role in combating the COVID-19 pandemic through a variety of appealing applications, ranging from outbreak tracking, virus detection to treatment and diagnosis support. On the one hand, AI is able to provide viable solutions for fighting the COVID- 19 pandemic in several ways. For example, AI has proved very useful for supporting outbreak prediction, coronavirus detection as well as infodemiology and infoveillance by leveraging learning-based techniques such as ML and DL from COVID-19-centric modeling, classification, and estimation. Moreover, AI has emerged as an attractive tool for facilitating vaccine and drug manufacturing. By using the datasets provided by healthcare organizations, governments, clinical labs and patients, AI leverages intelligent analytic tools for predicting effective and safe vaccine/drug against COVID-19, which would be beneficial from both the economic and scientific perspectives. On the other hand, big data has been proved its capability to tackle the COVID-19 pandemic.  Big data potentially provides various promising solutions to help fight the COVID-19 pandemic. By combining with AI analytics, big data helps us to understand the COVID-19 in terms of virus structure and disease development. Big data can help healthcare providers in various medical operations from early diagnosis, disease analysis to prediction of treatment outcomes. With its great potentials, the integration of AI and big data can be the key enabler for governments in fighting the potential COVID-19 outbreak in the future. 

Some recommendations can be considered to promote the COVID-19 fighting. First, AI and big data-based algorithms should be optimized further to enhance the accuracy and reliability of the data analytics for better COVID-19 diagnosis and treatment. Second, AI and big data can be incorporated with other emerging technologies to offer newly effective solutions for fighting COVID-19. For example, data analytic tools from Oracle cloud computing have been leveraged to design Vaccine, a new vaccine candidate against the COVID-19 virus \cite{200}. \textcolor{black}{More interestingly, recent studies have showed that 5G wireless technologies (e.g., drones, IoT, localization) can be utilized to combat the COVID-19 pandemic \cite{chamola2020comprehensive} via a number of applications, such as delivery of testing samples, goods transport, social distancing, and people’s movement monitoring for outbreak tracking.} Final, non-technology measures such as social distancing restrictions 
still play a highly important role in slowing down the virus spread and thus need to be implemented effectively under the management of government agencies \cite{lewnard2020scientific}, aiming to control the COVID-19 pandemic in the future.

\section{Conclusion} 
\label{Sec:Conclusion}
In this paper, we have presented a survey on the state-of-the-art solutions in the battle against the COVID-19 pandemic. 
Firstly, we have provided an introduction of the COVID-19 virus, the fundamentals and motivations of AI and big data for finding fast and effective approaches that can effectively combat the COVID-19 disease. 
Then, we have reviewed the applications of AI for detection and diagnosis, tracking and predicting the outbreak, infodemiology and infoveillance, biomedicine and pharmacotherapy. The applications of big data for the COVID-19 disease have been also presented, including outbreak prediction, virus spread tracking, diagnosis and treatment, and vaccine and drug discovery. 
Furthermore, we have discussed the challenges needed to overcome for the success of AI and big data in fighting the COVID-19 pandemic. Finally, we have highlighted important lessons and recommendations for the authorities and research communities.


\bibliographystyle{IEEEtran}
\bibliography{Bibliography}


\begin{IEEEbiography}[{\includegraphics[width=1in,height=1.25in,clip,keepaspectratio]{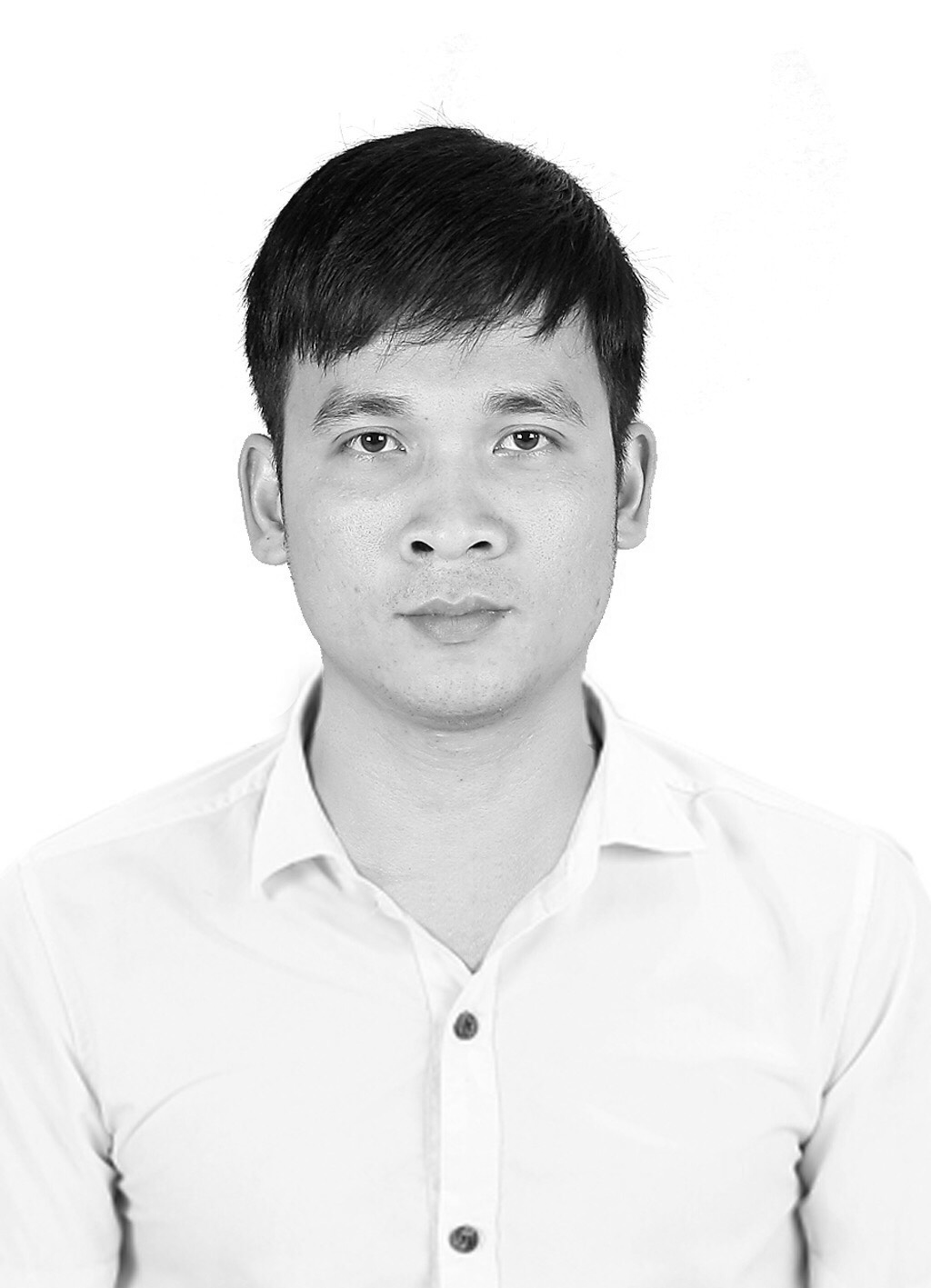}}]{Quoc-Viet Pham} (M'18) received the B.S. degree in electronics and telecommunications engineering from Hanoi University of Science and Technology, Vietnam, in 2013, and the Ph.D. degree in telecommunications engineering, from Inje University, South Korea, in 2017. He is currently a research professor at Research Institute of Computer, Information and Communication, Pusan National University, South Korea. From Sept. 2017 to Dec. 2019, he was with Kyung Hee University, Changwon National University, and Inje University on various academic positions. He received the best PhD dissertation award in Engineering from Inje University in 2017. His research interests include convex optimization, game theory, and machine learning to analyze and optimize edge/cloud computing, and 5G and beyond networks. He is a member of the IEEE. 
\end{IEEEbiography}

\begin{IEEEbiography}[{\includegraphics[width=1in,height=1.25in,clip,keepaspectratio]{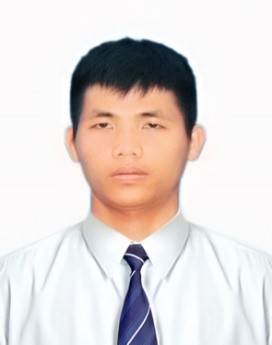}}]{Dinh C. Nguyen} received the B.S. degree (first class honors) in electrical and electronic engineering from Ho Chi Minh City University of Technology, Vietnam, in 2015. Currently, he is a Ph.D. scholar in the School of Engineering, Deakin University, Victoria, Australia. His research interests focus on security and privacy in Internet of Things (IoT), mobile cloud computing, and blockchain. He is currently working on adopting blockchain for secure communication networks including clouds, IoT, and healthcare.  He is a recipient of the Data61 PhD scholarship, CSIRO, Australia. 
\end{IEEEbiography}

\begin{IEEEbiography}[{\includegraphics[width=1in,height=1.25in,clip,keepaspectratio]{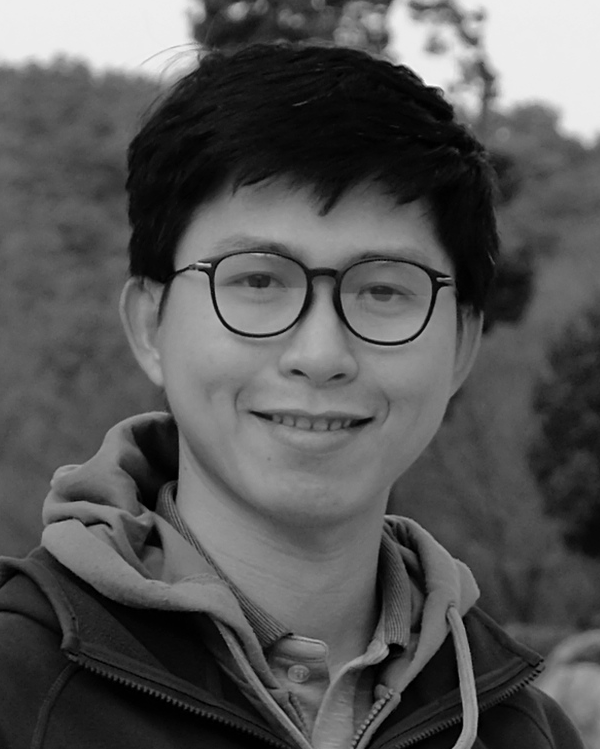}}]{Thien Huynh-The}(S'15, M'19) 
	received the B.S. degree in electronics and telecommunication engineering
	from Ho Chi Minh City University of Technology and Education, Vietnam, in 2011,
	and the Ph.D. degree in computer science and engineering from Kyung Hee University (KHU), South Korea, in 2018. He is awarded with the Superior Thesis Prize by KHU. 
	
	He is currently a Postdoctoral Research Fellow with ICT Convergence Research Center at Kumoh National Institute of Technology, South Korea. His current research interest includes radio signal processing, digital image processing, computer vision, machine learning, and deep learning.
\end{IEEEbiography}

\begin{IEEEbiography}[{\includegraphics[width=1in,height=1.25in,clip,keepaspectratio]{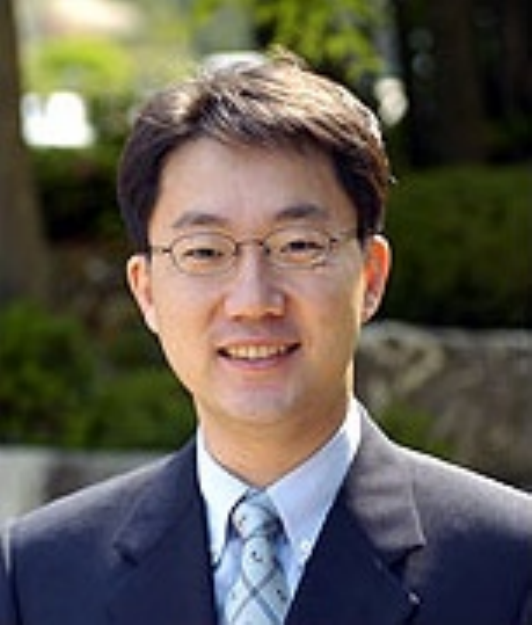}}]{Won-Joo Hwang} (S'01-M'03-SM'17) received the B.S. and M.S. degrees in computer engineering from Pusan National University, Busan, South Korea, in 1998 and 2000, respectively. He received the Ph.D. degree in information systems engineering from Osaka University, Osaka, Japan, in 2002. From 2002 to 2019, he was a Full Professor at the Inje University, Gimhae, South Korea. Currently, he is a Full Professor in the Biomedical Convergence Engineering Department at the Pusan National University. His research interests include optimization theory, game theory, machine learning and data science for wireless communications and networking. He is a senior member of the IEEE.
\end{IEEEbiography}

\begin{IEEEbiography}[{\includegraphics[width=1in,height=1.25in,clip,keepaspectratio]{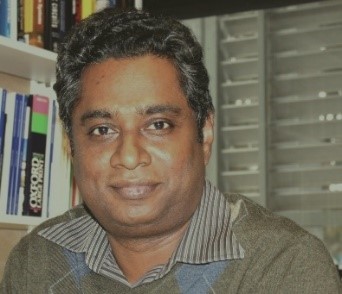}}]{Pubudu N. Pathirana} was born in 1970 in Matara, Sri Lanka, and was educated at RoyalCollege Colombo. He received the B.E. degree (first class honors) in electrical engineering and the B.Sc. degree in mathematics in 1996, and the Ph.D. degree in electrical engineering in 2000 from the University ofWestern Australia, all sponsored by the government of Australia on EMSS and IPRS scholarships, respectively. He was a Postdoctoral Research Fellow at Oxford University, Oxford, a Research Fellow at the School of Electrical Engineering and Telecommunications, University of New South Wales, Sydney, Australia, and a Consultant to the Defence Science and Technology Organization (DSTO), Australia, in 2002. He was a visiting associate professor at Yale University in 2009. Currently, he is a full Professor and the Director of Networked Sensing and Control group at the School of Engineering, Deakin University, Geelong, Australia and his current research interests include Bio-Medical assistive device design, human motion capture, mobile/wireless networks, rehabilitation robotics and radar array signal processing. He is a senior member of the IEEE.
\end{IEEEbiography}

\EOD

\end{document}